\documentclass{ws-bme}
\usepackage{multicol,balance}
\usepackage{url}
\usepackage{cite}
\usepackage{hyperref}
\usepackage{booktabs} %
\hypersetup{
    colorlinks=true,
    linkcolor=blue,
    filecolor=magenta,      
    urlcolor=cyan,
    pdftitle={Overleaf Example},
    pdfpagemode=FullScreen,
    }
\usepackage{graphics}
\usepackage{graphicx}
\usepackage[caption=false,font=footnotesize]{subfig}
\begin{document}

\markboth{Beghoura et al}{An Improved CovidConvLSTM model for pneumonia-COVID-19 detection and classification}

\title{An improved CovidConvLSTM model for pneumonia-COVID-19 detection and classification}
\author{Imane Beghoura $^{a}$, Mustapha Benssalah $^{a}$ and Fazia Sbargoud $^{b}$ }

\address{$^{a}\!$Signal Processing Laboratory, Ecole Militaire Polytechnique (EMP), Algiers, Algeria\\
\email{Imane.beghoura@emp.mdn.dz}, 
\email{Mustapha.benssalah@emp.mdn.dz }}
\address{$^{b}\!$Complexe Systems Control and Simulators Laboratory, EMP, Algiers, Algeria\\
\email{Fazia.sbargoud@emp.mdn.dz}}

\maketitle

\footnotetext{$^{\ddag}$Corresponding author: Imane Beghoura, Signal Processing Laboratory, Ecole Militaire Polytechnique, Algiers, Algeria. E-mail: Imane.beghoura@emp.mdn.dz}

\begin{abstract}

Recently, COVID-19 pandemic has rapidly evolved into a critical global health crisis, profoundly impacting daily life. As a result, Computer-Aided Diagnosis (CAD) systems have gained significant interest for its massive computational capabilities, which facilitate the rapid analysis and interpretation of large amounts of medical imaging data, crucial for managing the high influx of COVID-19 cases. In particular, Deep Learning (DL) techniques have emerged as critical tools to assist radiologists and pulmonologists in distinguishing COVID-19 patients from other pneumonia types and healthy cases. Unfortunately, existing DL techniques face several challenges such as overfitting, performance degradation, feature irrelevance and redundancy, vanishing gradient problem, and high computational complexity. In this paper we address these challenges by introducing an enhanced Convolutional Neural Network (CNN) algorithm that combines bottleneck based model RegNetX002, ConvLstm layer, and Squeeze and Excitation (SE) block. Specifically, the RegNetX002 and the ConvLstm layer are used for features map extraction and feature quality enhancement, while the attention mechanism SE block is employed to improve feature representation by highlighting important channel features and suppressing unimportant features. More importantly, The bottleneck module facilitates the extraction of more abstract features while lowering computational costs. Additionally, it incorporates residual connections that helps reducing the vanishing gradient problem. Balanced CPN-CXRPA dataset and imbalanced CXRI-P/C-CXR dataset are used to assess the proposed model. Performance metrics such as accuracy and F1 score are used to evaluate the model’s efficiency. Using the CPN-CXRPA dataset, our model achieved an accuracy of 98.22\%. For the CXRI-P/C-CXR dataset, it achieved 98.78\% of both accuracy and F1 score. The experimental results show that this framework outperforms existing models in terms of performance and computational complexity, making it a promising tool for managing this health crisis effectively and efficiently.

\end{abstract}



\begin{multicols}{2}

\section*{INTRODUCTION}

Pneumonia is an acute respiratory infection affecting the lungs. It is one of the leading infectious causes of mortality in children worldwide. In 2019, pneumonia caused the death of 740180 children, accounting for 14\%  of all fatalities in children under five~\cite{1}. Furthermore, according to the World Health Organization estimate, two-thirds of the world’s population do not have access to radiologists, which can reduce the detection of the disease in areas with medical deserts or developing countries~\cite{1}.
By December 2019, COVID-19 appeared. Characterized by the rapidity of its spreading, this viral pneumonia become a public health emergency of international concern~\cite{3}. By April, 2020, more than 84000 cases were confirmed in China and more than 2.30 million cases globally~\cite{4}. Among the various available diagnostic methods, Reverse Transcription Polymerase Chain Reaction (RT-PCR) has been the most commonly used technique for detecting the virus. However, it is less sensitive than chest imaging, including Chest x-ray (CXR) scanning~\cite{6}, especially for the early diagnosis and treatment of COVID-19 patients. Moreover, the process of collecting and processing RT-PCR bulk samples can take days before receiving the results of a single sample~\cite{7}. Compared to RT-PCR, chest imaging is a more reliable, valuable, and rapid technology for the early detection and evaluation of this pathology, especially in epidemic areas. Almost all hospitals have X-ray chest scanners and Computerized Tomography (CT) screening systems. However, the diagnosis of COVID-19 based on chest images requires a radiology expert, which can be laborious and time-consuming. Additionally, visual examination of these images might lead to an inaccurate diagnosis due to the inter-/intra- personal observation.

To overcome these barriers, Computer-Aided Diagnosis (CAD) systems have emerged as significant tools for fast process, analyse and interpretation of large amount of medical images which provides a valuable assistance for radiologists. Machine Learning (ML) algorithms have been increasingly proposed as solutions to enhance detection and classification tasks using CXR images or CT-scans~\cite{Santosh2021}. With improvements in image processing technology which enhance the images resolution, the need for effective computing models and image analysis methods has become more crucial.  
DL methods improved pneumonia detection efficiency and demonstrated high accuracy compared to previous ML approaches~\cite{siddiqi2024deep}. In particular, Recurrent Neural Networks (RNNs) and Convolutional Neural Networks (CNNs) have shown significant promise in outperforming conventional models.

RNNs, particularly Long Short-Term Memory (LSTM) networks, have been explored for medical images classification due to their ability to process sequential data~\cite{hochreiter1997long}~\cite{udriștoiu2021real}. While RNNs excel at capturing temporal dependencies, they face significant limitations. In fact, the vanishing gradient problem can hinder training, especially in deep networks, making it challenging to learn from long sequences. Additionally, RNNs are not inherently designed for spatial feature extraction, which is crucial in image classification tasks, limiting their effectiveness in medical imaging scenarios~\cite{shankar2021optimal}.

CNNs have emerged as the dominant architecture for image classification tasks~\cite{rajkumar2021covid}~\cite{sethy2020detection}, including pneumonia and COVID-19 detection. This domination can be attributed to the convolutional layers' filters within a CNN model, which enable the extraction of more accurate features from image data, thereby leading to superior classification performance.
 Various studies have demonstrated the effectiveness of CNN architectures in achieving high accuracy in classifying chest X-ray images. For instance, Zhang et al~\cite{zhang2020covid}. applied ResNet18 and achieved a 95.18\% accuracy in binary classification of COVID-19. In another exploration, Wang et al.~\cite{16} proposed COVID-Net, an open-source deep neural network for COVID-19 that achieved 92.4\% accuracy in classifying standard non-COVID and COVID-19 pneumonia classes. To the best of our knowledge, COVID-Net is one of the first open-source networks designed for COVID-19 detection from CXR images. Habib et al.~\cite{20} introduced CheXNet, a deep CNN model, to accurately detect pneumonia from chest X-ray images using a 121-layer DenseNet architecture.
Although CNNs yield accurate results, they have tendency to overfit on small datasets and a requirement for substantial computational resources during training. Furthermore, CNNs primarily focus on spatial features and may not sufficiently address temporal changes in sequential imaging data, which can be a critical factor in diagnosing diseases that progress over time~\cite{wang2023pneunet}.

To leverage the strengths of both CNNs and RNNs, researchers have explored hybrid models that combine these architectures. By using CNNs for spatial feature extraction and RNNs for temporal analysis, these models can potentially improve classification performance. In this context, Naeem et al.~\cite{naeem2021cnn} introduced a hybrid CNN–LSTM model integrated with a multi-level feature extraction technique that used GIST and scale-invariant feature transform to simplify CNN training. This approach significantly enhanced the accuracy of both detection and severity classification of COVID-19 using CT and CXR images. It demonstrated high performance of 98.94\% accuracy in COVID-19 detection and 83.03\% accuracy in severity classification. Islam et al.~\cite{islam2020combined} developed a CNN-LSTM model for COVID-19 detection and classification. This model presented exceptionally high results of of 99.4\% accuracy and 99.3\% recall. However, this approach can lead to increased model complexity and longer training times, which may not always yield significant performance gains. The integration of CNN and RNN also increase the model architecture complexity~\cite{kanjanasurat2023cnn}, making it more challenging to optimize and deploy in clinical settings.

In contrast to traditional RNNs, the combination of CNNs with Convolutional Long Short-Term Memory (ConvLSTM) layers offers a more effective solution for spatiotemporal data~\cite{33}. ConvLSTM is designed to capture both spatial and temporal dependencies, making it particularly suitable for classification tasks such as COVID-19 detection from sequences of medical images. This hybrid approach has shown improved performance in various studies, as it retains spatial hierarchies while effectively modeling the temporal evolution of the data. Also, ConvLSTM can lead to faster convergence and better generalization on unseen data. Although, it may still introduce additional computational complexity.

Recent research has also explored the use of Ensemble Transfer Learning (ETL) methods for enhancing COVID-19 detection~\cite{gifani2021automated,kaleem2023ensemble,dubey2023ensemble,roy2022early,hariri2023advanced}. ETL combines multiple pre-trained models to improve classification performance. This approach has been particularly effective for COVID-19 and pneumonia classification, as it leverages the strengths of various architectures to enhance accuracy and robustness. For instance, two different set transfer learning systems were suggested by Gianchandani et al.~\cite{27} for COVID-19 diagnostic using CXR pictures. They detected COVID-19, bacterial pneumonia, and other pathogens and improved detection performance using pre-trained models. Another model by Singh et al.~\cite{29} combined ResNet152V2 and VGG16 for accurate classification of chest X-ray scans into COVID-19, pneumonia, tuberculosis, and healthy cases. Gianchandani et al.~\cite{27} also proposed a modified VGG16 and DenseNet201 with ResNet152V2 for multiclass and binary classification. Subhrajit et al.~\cite{31} employed a fuzzy ensemble model using pre-trained CNNs for COVID-19 detection from chest X-ray images. The model was called CovidConvLSTM and showed relevant results, achieving a high accuracy of 98.62\%. While ensemble methods have reported accuracies exceeding 98\%~\cite{hussein2023lightweight}, their main limitation is their computational complexity, as they require significant resources for training and inference due to the need for simultaneous process of multiple models.

Moreover, existing COVID-19 and pneumonia diagnosis models suffer from overfitting, performance degradation, irrelevance and redundancy of features and vanishing gradient problem in addition to computational complexity. Previous works in this domain have been limited especialy by the insufficient amount of available data for training, which can hinder the ability of models to learn accurate features and achieve optimal performance levels. To deal with the above-mentioned issues, the main goal of this contribution is to suggest a robust DL model for the automated and rapid COVID-19 and pneumonia diagnosis.

Recently, studies have shown that bottleneck-based models can offer a more efficient alternative, particularly in scenarios where computational resources are limited~\cite{zhou2020rethinking,hoang2022rethinking}. Since bottleneck architectures reduce the dimensionality of input features, they have been used to maintain model performance while significantly lowering computational costs~\cite{he2016deep,sandler2018mobilenetv2}. These models simplify the learning process by focusing on important features, making them a compelling option to optimize the model performance. Additionally, the bottleneck approach addresses the limitations of ETL by providing a balance between accuracy and efficiency, which is critical for the rapid deployment of diagnostic tools in healthcare settings.

In this paper, we introduce an efficient classification model for COVID-19 and other pneumonia infections inspired by CovidConvLSTM architecture and based on a bottleneck module. The proposed model inherited the best aspects of the original model~\cite{31} CovidConvLSTM and improved it further.

\subsection*{Motivation and major contributions}


Our model is designed to reduce overfitting through regularization techniques and model architecture. It ensures high performance by addressing class imbalance and optimizing hyperparameters. Feature selection using the RegNet model, ConvLSTM layer and a SE block is applied to eliminate irrelevant and redundant features, enhancing the model's effectiveness. Additionally, the proposed model is optimized for computational efficiency, balancing model complexity and resource requirements by using a bottleneck based model to ensure rapid and reliable diagnostics. This comprehensive approach aims to overcome the limitations of existing methods and advance the field of medical diagnostics using deep learning.

The objectives and significant contributions of this research are outlined as follows:

\begin{itemize}
  \item We introduce a DL model that incorporates linear bottlenecks which enhance the efficiency of COVID-19 diagnosis and reduces computational complexity.
  \item We apply a ConvLstm layer to the feature maps to improve the image representation and the inter-dependencies among feature maps, which helps our model to discern spatial dependencies.
  \item We incorporate an attention mechanism through SE block to enhance the model performance by highlighting important features and removing non-important features.
  \item We conduct two experiments on balanced and unbalanced datasets to asses and compare the performance of our suggested model with existed CNN models and CovidConvLSTM method. 
\end{itemize}

\section*{MATERIALS AND METHODS}

This section describes the methodologies used to develop and train our model, aimed at enhancing the diagnosis of COVID-19 and other pulmonary conditions using CXR images. It provides a comprehensive description of the datasets used for training, validation and test, the architecture of the model and the training hyper-parameters. 

\subsection*{Datasets description and pre-processing }

The proposed model is evaluated using two datasets. The first one, covid19-pneumonia-normal-chest-xraypa-dataset (CPN-CXRPA), is a balanced dataset from Kaggle~\cite{35}. It consists of X-Ray scans of individuals with COVID-19 positive, COVID-19 negative and pneumonia cases. It comprises 4575 samples distributed in three classes: normal (1525), Covid-19 (1525), and pneumonia (1525). The dataset is divided into 3111 training samples, 732 test samples, and 732 validation samples.
The second dataset is a hybrid compilation of two publicly accessible datasets: chest-xray pneumonia (CXRI-P)~\cite{36} and covid-chest x-ray-dataset (C-CXR)~\cite{37}, collected from Kaggle~\cite{38} and GitHub~\cite{39}, respectively.  
The CXRI-P dataset contains images from retrospective cohorts of pediatric patients aged from one to five years from China's Guangzhou Women and Children's Medical Center, while the C-CXR dataset consists of CXR and CT scan images from COVID-19 patients. For this study, only chest X-rays are considered. This hybrid dataset contains a total of 6140 samples, distributed across three classes: normal (1341 images), COVID-19 (924 images), and pneumonia (3875 images). It is divided into 4420 training samples, 1229 test samples, and 491 validation samples. Both datasets provide a comprehensive foundation for evaluating the effectiveness and robustness of the proposed deep learning model in detecting and classifying lung disorders into normal, COVID-19, and pneumonia categories. The balanced distribution and substantial sample sizes ensure that the model can be trained and tested rigorously, contributing to its potential in assisting medical professionals in diagnosing and managing respiratory conditions.



In this work, X-ray images are resized to (224, 224 ,3). It is common practice to resize images to (224, 224) in CNN models for image classification due to the architecture of popular image classification models which are developed and trained on this image size to extract relevant features while keeping the number of parameters manageable, thereby avoiding overfitting.

Data splitting into training, validation, and testing is a crucial step. In this work, we divide the first dataset into 68\% for training, 16\% for validation, and 16\% for testing, and the second dataset into 72\% for training, 8\% for validation, and 20\% for testing.

\subsection*{Proposed architecture}

The suggested method consists of an automated decision tool for COVID-19 and pneumonia detection from CXR images. The model was inspired by CovidConvLSTM architecture, where the authors combined three different CNN models using Sugeno fuzzy integral~\cite{31}. To strengthen the feature extraction and enhance their representation, the authors add a ConvLSTM layer and an attention mechanism to each model. 

\begin{figure*}
\centering
\includegraphics[width=2\columnwidth]{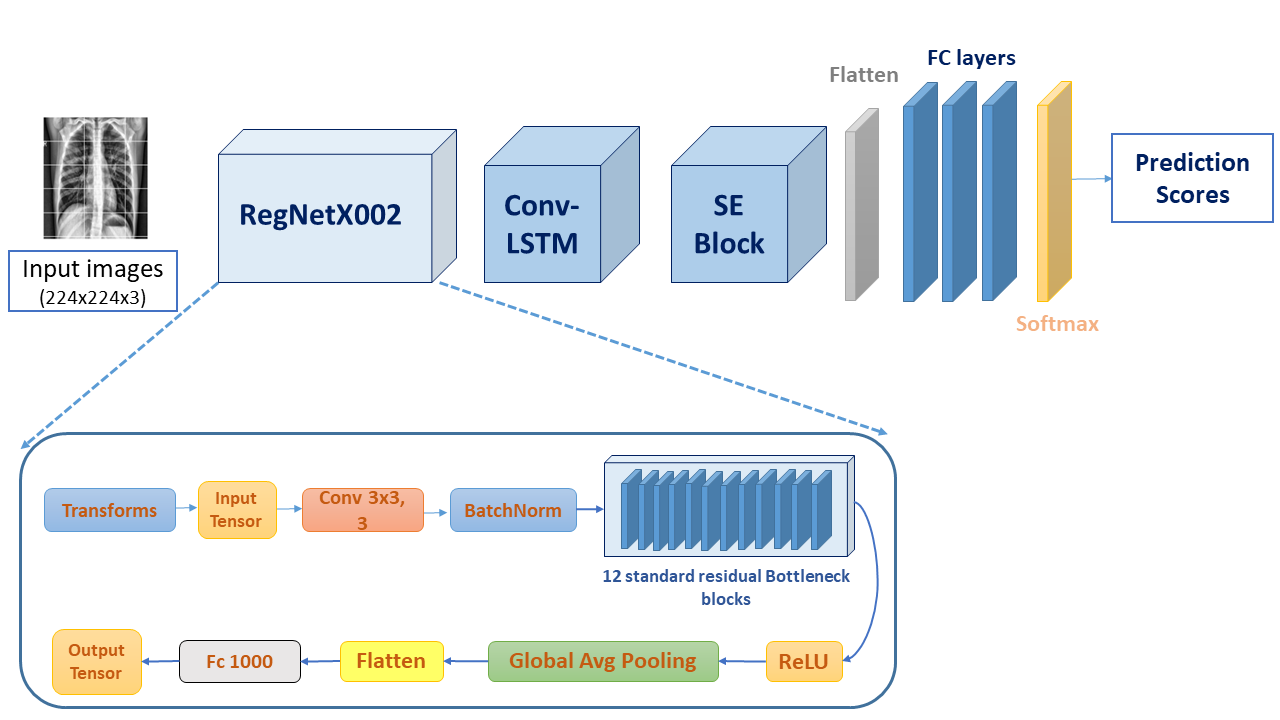}
\caption{Proposed model architecture featuring RegNetX002 block.}
\label{fig1}
\end{figure*}

The main aim of this study is to leverage the efficiency of CovidConvLSTM and improve it while concurrently refine the complexity of its architecture. This is achieved through the suppression of the aggregation and ensemble learning technique and the use of a linear bottleneck model instead. As it is shown in Figure~\ref{fig1}, our technique is based on the RegNetX002 model proposed by Ilija Radosavovic et al.~\cite{32}, followed by a ConvLSTM layer with 512 filters to enable establishing correlated features. The features are compressed, and the parameters are adjusted using a squeeze and excitation (SE) block, which highlights the crucial features and suppresses redundant or non-important features. Three fully connected layers are added for the classification task.

The suggested model incorporates the best aspects of the CovidConvLSTM and the RegNet model. Evaluated on two publicly available datasets.

\subsubsection*{RegNetX002}

The RegNet is a network design space that is founded on the standard residual bottleneck block with group convolution, which is intended for image classification tasks. It was introduced by Radosavac et al.~\cite{32} as part of a broader examination of scalable and effective convolutional neural networks (CNNs). The authors presented a methodical strategy for designing CNNs and assessing the efficiency of RegNet on various image classification benchmarks. It was constructed to resolve the difficulties of scaling up CNNs to handle high-resolution images while retaining high accuracy. It employs residual connections and shortcuts to enable easier optimization and faster convergence. In addition, it features a lightweight design with fewer parameters, which makes it more suitable for deployment on edge devices and other resource-limited environments.

RegNet Network structure can be determined by six parameters: d (depth), $w_0$ (initial width), $w_a$ (slope), $w_m$ (width multiplication factor), b (bottleneck), and g (group width). These parameters are used in Eqs. \eqref{eq1}, \eqref{eq2} and \eqref{eq3} to compute the depths of the blocks in the network and to generate the block width $u_j$ for each block $j<d$, based on the linear parameterization of blocks widths. The resulting design space is referred to as RegNet, and each model is obtained by setting these parameters to address different properties.

\begin{equation}
u_j = w_0 + w_aj \hspace{1cm}  \rm{for} \hspace{1cm}   0<j<d
\label{eq1}
\end{equation}
\begin{equation}
u_j=w_0w_m^{s_j}
\label{eq2}
\end{equation}
\begin{equation}
w_j=w_0w_m^{[s_j]} 
\label{eq3}
\end{equation}

The RegNetX002 model is structured into several stages, each of which is made up of multiple blocks. These blocks collectively form the stem that takes the image input, the body that handles the majority of computations, and the Head that carries out the classification task. The type of block used is the standard residual bottleneck block with group convolution.

The bottleneck block is known for its efficiency in reducing computational complexity while maintaining high performance. It uses a three-layer architecture that compresses and then expands the feature dimensions, effectively bottlenecking the information flow in the middle layer. The first layer aims to reduce the dimensionality of the input features, which helps decreasing the computational load. The second layer performs the main convolution operation on the reduced number of channels. The last convolutional layer intends to restore the reduced feature maps by expanding the feature maps dimensionality. This architecture incorporate a skip connections (identity mappings) that helps in reducing the vanishing gradient problem. 


\subsubsection*{LSTM convolutional layer}

Convolutional LSTM is a variant of LSTM  that combines the strengths of CNNs and LSTMs. ConvLSTM layers have been widely adopted in various applications, including computer vision, video analysis, natural language processing, and time series prediction, due to their ability to capture spatial and temporal dependencies in data. The concept was first introduced by Shi et al. in their groundbreaking paper in 2015~\cite{33}. The ConvLSTM layers architecture builds upon the traditional LSTM structure by incorporating convolutional operations alongside recurrent operations. It comprises four essential components: the input gate $i_t$, the forget gate $f_t$, the output gate $o_t$, and the cell state $c_t$. These components are updated at each time step and collectively determine the flow of information through the layer. This allows ConvLSTM to represent both spatial and temporal dependencies simultaneously, making it particularly suitable for tasks that involve processing data with spatial structures, such as images and videos. 


\subsubsection*{Squeeze and Excitation block}

The SE block is an attention mechanism and a neural network architecture that was originally introduced by Hu et al. in 2018~\cite{34}. The SE block is commonly employed in CNNs to improve feature representation by re-calibrating the channel-wise feature responses of a CNN. It typically consists of two primary operations: a squeeze operation that reduces the spatial dimensions of a feature map and an excitation operation that involves learning a set of channel-wise weights to highlight informative features.



The feature maps of the CNN model and the block SE input pass through global average pooling as an initial step to reduce their spatial dimensions. This reduction process is referred to as "squeezing". Then the resulting 1D feature maps undergo an excitation operation, where a set of channel-wise weights is learned. These weights are acquired during training and facilitate the network to focus on different areas of the feature maps for various tasks. Finally, the output of the SE block is added back to the original feature maps to regulate their responses. SE blocks have demonstrated remarkable performance improvements in CNNs, particularly in image classification and object detection applications. These blocks are frequently used in residual networks and are incorporated into many widely-used architectures.

\subsubsection*{Hyperparameters }

The proposed model has been trained in the current set of experiments using a set of optimized hyperparameters. To prevent the overfitting during the model training and the weights updating, which have been initiated with ImageNet dataset, the number of epochs has been set to 10 and the learning rate has been set to 4e-4. The optimizer used in this study is Adam and the Rectified Linear Unit (ReLU) has been used as the activation function.
In this work, the RegNetX002 is fine-tuned on the target set to generate feature maps and combined with ConvLSTM layers with 512 filters and a kernel size of 1x1. A SE block has been used to focus on the feature maps, where the value of the ratio, or reduction ratio, has been set to 16. The selection of a suitable ratio number for a SE block is contingent upon the particular task, the dataset under consideration, the size of the input, and the balance model complexity and performance. It is commonplace to employ a default value of 16 for the ratio number to ascertain an optimal number of channels in the squeezed feature map, which corresponds to one-sixteenth of the number of channels present in the original feature map. Following this step, three fully connected layers, each with 4096 nodes, have been employed. Finally, a softmax layer of size 3×1 has been applied to classify the images into three categories by using the extracted features. The categorical cross-entropy has been chosen as the loss function during the classification process.
To reduce the common stochastic behavior of CNNs that can lead to varying accuracy values within a specific range, we used a seed value of 3 to initialize the random number generator, ensuring that the same sequence of random numbers is generated each time the model is trained.

\subsubsection*{Computational Complexity Evaluation}

In our quest to develop an advanced neural network model for pneumonia and COVID-19 detection and classification, it is essential to address the balance between complexity and performance. while building upon the foundations of state-of-the-art models, our proposed model introduces key innovations that optimize diagnostic accuracy while substantially reducing computational cost. A vital aspect of this comparison lies in the evaluation of both FLOPS (Floating-Point Operations Per Second) and hyperparameters, metrics that provide valuable insights into computational efficiency and feasibility.

The evaluation reveals a significant divergence between our proposed model and the CovidConvLstm model. While our model outperformed the latter, achieving slightly higher diagnostic accuracy, it simultaneously demonstrated significantly lower computational complexity of 1,882,592,528 FLOPs compared to the 46,703,177,240 FLOPs required by the CovidConvLSTM model, representing a 95.97\% reduction. Additionally, the proposed model utilizes 140,515,747 parameters, which is notably lower than the 363,996,809 parameters used by the CovidConvLSTM model, resulting in a 61.40\% reduction. Specifically, our model exhibited a notable reduction in FLOPS, reflecting its efficiency in terms of computational cost. This advantage arises from the architectural design of the Regnet network that leverages more efficient layers and well allocated parameters, a departure from the excessive complexity seen in some state-of-the-art models.


Furthermore, the hyper-parameter count of our proposed model is notably lower, reflecting a meticulous approach to parameter management, which focuses on optimizing model performance without introducing irrelevant complexity. This reduction in hyper-parameters proves the efficiency of our architectural enhancements.

\section*{RESULTS}

This section outlines the experimental results and performance evaluation of the proposed model, which was implemented in Python with keras and tensorflow frameworks and tested on two publicly available lung disease datasets. The purpose of both datasets is to differentiate COVID-19 cases from other pneumonia cases and normal cases. To evaluate the efficacy of the proposed model, we conducte a comparative study with similar models based on metrics such as overall accuracy and computational complexity. In addition, we analyzed the proposed model using various performance metrics, including F1 score, recall, and precision. To gain insight into the nature of errors in the model's predictions, we employed a confusion matrix to allocate predictions to the true ground classes.

\subsection*{Experiments on balanced Dataset}

The efficacy of our proposed model for the CPN-CXRPA dataset is proven by the experimental outcomes of our method. Specifically, the accuracy of our improved model is reported to be 98.22\%, which is notably higher than the 97.27\% accuracy achieved by the CovidConvLSTM model and the significantly lower 64.48\% accuracy reached by RegNetX002 model. This high accuracy indicates that the proposed model is more effective in classifying images into the three categories: Covid-19, normal, and pneumonia on a balanced dataset.

Similarly, the f1-score, recall, and precision for our model are 98.22\%, which surpasses the 97.26\% F1 score, recall, and 97.29\% precision of the CovidConvLSTM model, and the lower scores reached by the RegNetX002 model, which are 64.19\% for the F1 score, 64.48\% for recall, and 64.89\% for precision.

Based on the provided results, the performance metrics superiority of the proposed model demonstrates advanced ability in classifying chest X-ray images with high reliability. The marked improvements over the CovidConvLSTM and RegNetX002 models underscore the effectiveness of our proposed model's architecture and methodology in addressing the classification challenges inherent in the CPN-CXRPA dataset.


The quantitative assessments and observations of the dataset are presented in the confusion matrix tables given in Figure~\ref{fig6}, which include true predictions and false predictions for each class. Analyzing the confusion matrices for our proposed model and the CovidConvLSTM model, it is notable that our model performed consistently across all three classes with high precision and recall. On the other hand, the CovidConvLSTM model appears to have performed efficiently for pneumonia cases but struggles with COVID-19 and normal cases.


\begin{figurehere}
\centering
\includegraphics[width = \columnwidth]{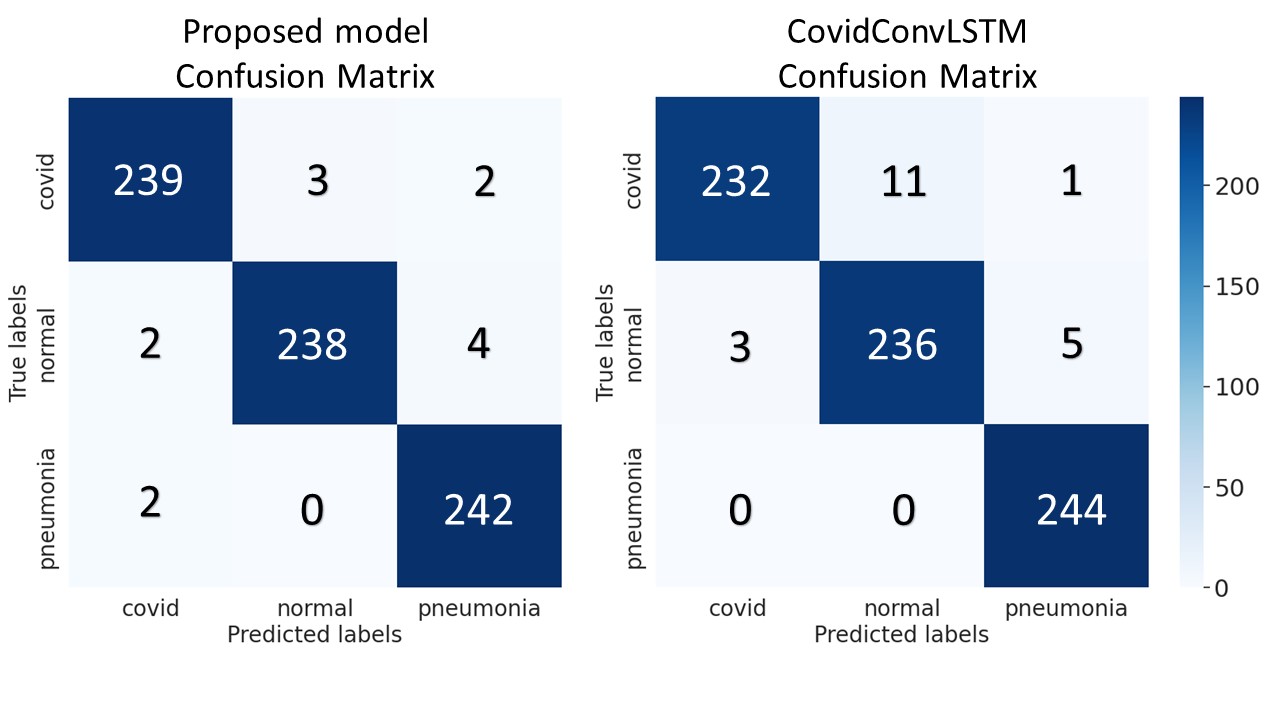}
\caption{Confusion matrices generated by our proposed model and CovidConvLSTM model both on CPN-CXRPA dataset.} 
\label{fig6}
\end{figurehere}

\begin{figure*}
	\centering
	\subfloat[CPN-CXRPA test set]{
		\includegraphics[width=0.48\textwidth]{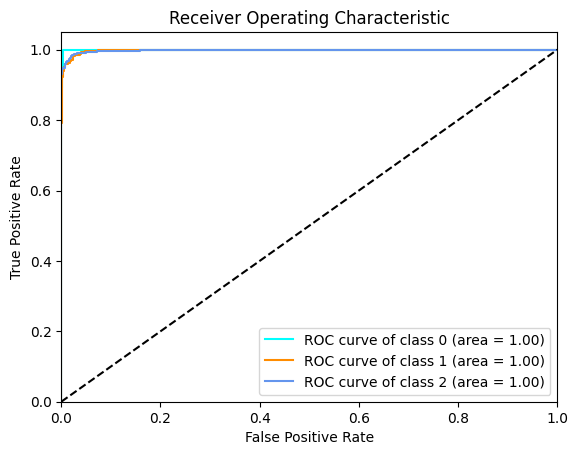}
		\label{fig88}}
	\subfloat[CXRI-P/C-CXR test set]{
		\includegraphics[width=0.48\textwidth]{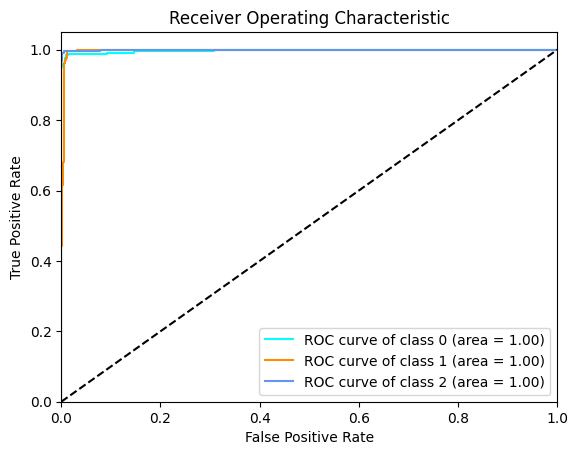}
		\label{fig89}}
	\caption{ROC curve of the proposed model.}
	\label{fig8}
\end{figure*}

ROC (Receiver Operating Characteristic) curves are a widely used tool for assessing the performance of models, particularly in the context of medical imaging applications. In the current study, ROC curves were employed to evaluate the effectiveness of the proposed model. As depicted in Figure~\ref{fig88}, the ROC curve for the proposed model demonstrates an Area Under the Curve (AUC) of 1 for each class. This exceptionally high AUC score signifies the model's capability in accurately diagnosing COVID-19 and pneumonia cases from chest X-ray (CXR) images, reflecting its substantial diagnostic accuracy and reliability.


\begin{tablehere}
\tbl{Performance comparison of different state-of-the-art models and the proposed model on CPN-CXRPA dataset }
{\begin{tabular}{@{}lcccr@{}}
\toprule
Model & Precision & Recall & F1-score & Accuracy\\
\colrule
RegNetX002 ~\cite{32} &  $67.19$ & $58.89$ & $62.50$ & $59.89$\\
CoroNet ~\cite{khan2020coronet}  & $	90.15$& $87.72$& $88.92$& $88.92$\\
NASNet-Mobile~\cite{naskinova2023transfer}  & $97.39$& $97.23$& $/$& $	96.02$\\
CovidConvLSTM ~\cite{31} & $97.29$ & $97.27$& $97.26$& $97.27$\\
\toprule
Proposed model & $98.23$ & $98.22$ & $98.22$ & $98.22$ \\
\botrule
\end{tabular}
\label{table4}
}
\end{tablehere}

Table~\ref{table4} provides a comprehensive comparison of five different models, including our proposed model, with respect to multiple key evaluation metrics. The table reveals the quantitative performance of these models, signifying their proficiency in classifying COVID, pneumonia, and normal cases.
Among the models scrutinized, our proposed model demonstrated exceptional performance across all metrics, which can be attributed to its high Precision of 98.23\%, a Recall of 98.22\%, an F1-score of 98.22\%, and an Accuracy of 98.22\%. These results clearly highlight the proposed model's effectiveness in accurately categorizing data into the three classes.

Overall, the results indicate that the improved RegNetX002 model is a promising approach for the classification of COVID-19, normal, and pneumonia cases. The use of RegNetX model combined with ConvLSTM layer, SE block, and fully connected layers have resulted in a significant improvement in model performance, achieving high accuracy, precision and recall across all three classes. These components are crucial in facilitating the model's ability to discern spatial dependencies and highlight essential features, while concurrently eliminating non-important information. Such advancements significantly refine the representation of images and elevate the classification performance.

\subsection*{Experiments on Imbalanced Dataset}

Similar to the initial dataset, we have evaluated the proposed model on the CXRI-P/C-CXR dataset, which is an imbalanced dataset. The findings of this evaluation demonstrate the performance of the suggested framework. Our model achieves an accuracy of 98.78\%, surpassing the 98.62\% accuracy of the CovidConvLSTM model and the considerably lower 59.89\% accuracy of the RegNetX002 model. Additionally, our proposed model demonstrates an f1-score of 98.78\%, outperforming the CovidConvLSTM’s f1 score of 98.67\% and the RegNetX002’s f1-score of 62.50\%. The recall and precision for the proposed model are 98.78\% and 98.78\%, respectively, which are higher than the recall of 98.67\% and precision of 98.34\% achieved by the CovidConvLSTM model, and significantly exceed the recall of 59.89\% and precision of 67.19\% of the RegNetX002 model. According to the obtained results, our model has shown a superior performance compared to the CovidConLSTM and the RegNetx002 models. The results denote that our implemented model has achieved a significant enhancement in metrics performance.


The CXRI-P/C-CXR dataset quantitative evaluations are illustrated in the Confusion Matrix Tables displayed in Figure~\ref{fig9}. Upon a thorough examination of the confusion matrices for the two models, our proposed model achieved notable precision in the normal and pneumonia classes, with minimal misclassifications. However, it exhibited some difficulty in accurately classifying instances in the COVID-19 class, resulting in two misclassifications. In contrast, the CovidConvLSTM model demonstrated strong performance in the COVID-19 class, with no misclassifications, showcasing its proficiency in distinguishing COVID-19 cases. Nevertheless, it experienced a higher number of misclassifications in the pneumonia class, with ten instances incorrectly classified as COVID-19.


\begin{figurehere}
\centering
\includegraphics[width = \columnwidth]{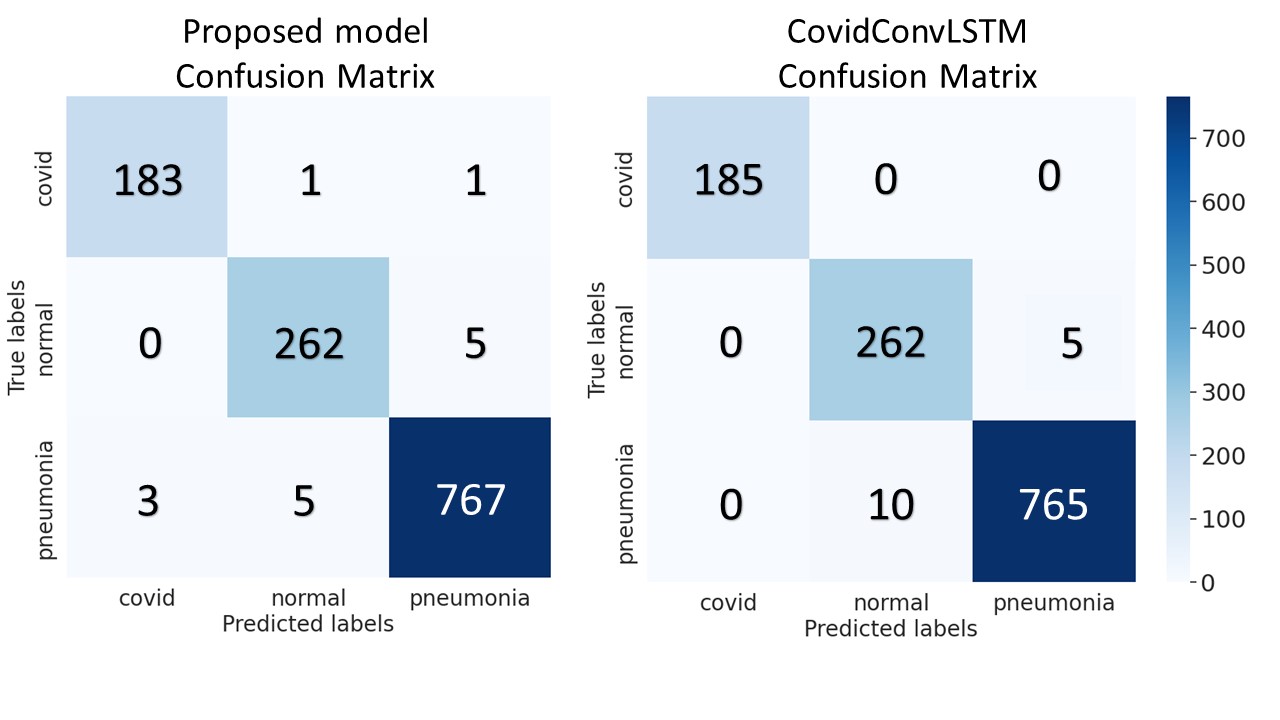}
\caption{Confusion matrices generated by our proposed model and CovidConvLSTM model on CXRI-P/C-CXR dataset.} 
\label{fig9}
\end{figurehere}


Both models show promise, but their strengths and weaknesses differ. The proposed model excels in normal and pneumonia class classification but faces challenges in COVID-19 classification. In contrast, the CovidConvLSTM model outperforms in COVID-19 class classification but struggles in pneumonia and normal classification.


Similar to the CPN-CXRPA dataset, Figure ~\ref{fig89} shows that the proposed model achieved a AUC of 1 for each class which signifies the model's ability to discriminate between the positive and the negative cases across all possible cases and indicate higher sensitivity and specificity.

\begin{tablehere}
\tbl{Performance comparison of different state-of-the-art models and the proposed model on CXRI-P/C-CXR dataset }
{\begin{tabular}{@{}lcccr@{}}
\toprule
Model & Precision & Recall & F1-score & Accuracy\\
\colrule
RegNetx002 ~\cite{32} &  $67.19$ & $58.89$ & $62.50$ & $59.89$\\
OptCoNet ~\cite{goel2021optconet} & $92.88$& $96.25$& $95.25$& $97.78$\\
Inverted Bell Ensemble ~\cite{paul2023inverted} & $97.21$& $97.81$& $97.50$& $97.97$\\
End to End CNN ~\cite{mukherjee2021deep} & $94.78$& $99.81$& $97.22$& $97.52$\\
CNN-LSTM model ~\cite{naeem2021cnn} & $95.00$ & $95.00$& $95.00$& $96.60$\\
F-RRN-LSTM model ~\cite{goyal2023detection} & $88.89$ & $95.41$& $92.03$& $94.31$\\
CovidConvLSTM ~\cite{31} & $98.34$ & $98.67$& $98.67$& $98.62$\\
\toprule
Proposed model & $98.78$ & $98.78$ & $98.78$ & $98.78$ \\
\botrule
\end{tabular}
\label{table5}
}
\end{tablehere}

Table ~\ref{table5} conducts a comprehensive comparative analysis between the proposed model and several other existing models. These models are evaluated with respect to their performance in classifying data into three distinct categories. Our proposed model exhibits outstanding overall performance when compared to the other models. It attains the highest precision, recall, f1-score, and accuracy, indicating its potential for excellence in the classification task.

\section*{DISCUSSION}

The present work explored the potential of integrating an attention mechanism and a ConvLSTM layer to a linear bottleneck based model in the detection of COVID-19 and pneumonia cases using CXR images. Our findings indicate that our improved framework can significantly enhance the rapid and cost-effective diagnosis of COVID-19 cases, as presented in different performance results. The ConvLSTM layer is used to learn spatial dependencies and to improve the interdependencies and image representations between feature maps, while the SE block is incorporated to improve attention and feature discriminability by highlighting crucial features and removing non-important features that may hinder the overall model performance.
Moreover, it should be noted that Subhrajit et al.~\cite{31} have proposed a framework that boasts a remarkable accuracy; however, an ensemble of transfer learning models is necessary to corroborate these results. This requirement not only increases the model's complexity but also intensifies its demand for extensive data sets, thereby presenting major limitations in terms of data availability in research contexts. In terms of model complexity, our approach stands out for its efficacy compared to the CovidConvLSTM model, while still achieving marginally superior performance. This significant finding suggests that incorporating linear bottlenecks model can enhance the efficiency of COVID-19 and pneumonia diagnosis and reduces computational complexity. Linear bottlenecks serve as a particular architectural feature within neural networks, focusing on reducing the dimensionality of the data at certain points in the network architecture. This approach effectively minimizes the number of parameters and computational complexity, thereby enhancing the model's overall efficiency.

In addition, we address the stochastic behavior of CNNs models, which can lead to varying accuracy values within a specific range. To minimize this, we used a seed value to initialize the random number generator. This practical strategy enhances the reproducibility of our results, potentially bolstering the reliability of our findings and reducing the variability in accuracy observed between different training runs.

\section*{CONCLUSION}

In this work, a novel COVID-19 and pneumonia infections DL-based detection model is introduced. The proposed framework is inspired by the CovidConvLSTM architecture. It inherited the best aspects of RegNet and CovidConvLSTM models and improved it further. The proposed model is based on a bottleneck based model RegNetX002, combined with a ConvLSTM layer to establish correlated features and capture spatial dependencies,  a SE block to highlight the crucial features and suppress non-important features, and three fully connected layers for classification tasks. 

The significance of our findings lies in the potential for DL frameworks to provide cost-effective solutions for COVID-19 diagnosis. Our study contributes to the growing body of literature on the use of deep learning algorithms in medical image analysis. It highlights the importance of developing accurate and efficient diagnostic tools for COVID-19. This can be achieved by analyzing various factors, such as the model complexity, the hyper-parameters and the dataset size. Moreover, the proposed framework takes advantage of the linear bottleneck to enhance the overall model efficiency and reduce computational complexity while improving the image representation and the features quality through the ConvLSTM layer and the SE block. The experimental results show that our model outperforms existing models in terms of performance. It achieves an accuracy of 98.22\% for CPN-CXRPA dataset and 98.78\% of both accuracy and F1 score for the CXRI-P/C-CXR dataset.
Our proposed framework not only achieves high performance and efficiency but also addresses computational challenges, making it an effective solution for real-world medical applications. 

\section*{ACKNOWLEDGMENTS}

I would like to express my sincere gratitude to Dr. Djeha Mohamed for his invaluable assistance in reviewing and providing insightful corrections to this article. His expertise and thoughtful feedback have significantly contributed to improving the quality of this work.


\bibliographystyle{ws-bme}
\bibliography{ws-bme}

\end{multicols}
\end{document}